\documentclass[pra,twocolumn,showpacs,floatfix]{revtex4}
\usepackage{graphicx}
\begin{document}

\title{Persistent currents in a two-component Bose-Einstein condensate
confined in a ring potential}
\author{J. Smyrnakis$^1$, M. Magiropoulos$^1$, Nikolaos K. Efremidis$^2$, 
and G. M. Kavoulakis$^1$}
\affiliation{$^1$Technological Education Institute of Crete, P.O. Box 1939, 
GR-71004, Heraklion, Greece
\\
$^2$Department of Applied Mathematics, University of Crete, GR-71004,
Heraklion, Greece}
\date{\today}

\begin{abstract}

We present variational and numerical solutions for the problem of stability 
of persistent currents in a two-component Bose-Einstein condensate of 
distinguishable atoms which rotate in a ring potential. We consider the 
general class of solutions of constant density in the two components 
separately, thus providing an alternative approach of the solution of the 
same problem given recently by Zhigang Wu and Eugene Zaremba [Phys. Rev. A 
{\bf 88}, 063640 (2013)]. Our approach provides a physically transparent 
solution of this delicate problem. Finally, we give a unified and simple 
picture of the lowest-energy state of the system for large values of the 
coupling. 

\end{abstract}
 \maketitle

\section{Introduction}

The problem of persistent currents in a toroidal/annular potential of 
Bose-Einstein condensed atoms (of a single species) has attracted a lot 
of attention in recent years. The experiments of Refs.\,\cite{Kurn, Olson, 
Phillips1, Foot, GK, Moulder, Ryu} have managed to create persistent 
currents in such trapping geometries and thus realize probably the most 
simple superfluid system that has been realized in the laboratory. These 
experiments are thus ideal for studying the fascinating effect of 
frictionless flow, which is one of the many problems associated with 
the more general effect of ``superfluidity".

A non-trivial extension of the problem of the stability of persistent 
currents is the one of mixtures of two distinguishable species. Remarkably, 
this problem has been realized and examined in the recent experiment of 
Ref.\,\cite{Zoran}, which makes its theoretical study even more interesting. 
In Ref.\,\cite{prl} it was shown that the stability of persistent currents 
is strongly affected by the addition of a second component. If $A$ and $B$ 
are the labels of the two species and ${\tilde \ell} = \ell \hbar = (L/N) 
\hbar$ is the angular momentum per particle, with $L \hbar = (L_A + L_B) 
\hbar$ being the total angular momentum and $N = N_A + N_B$ being the total 
population of the two species, it was shown that in the range $0 \le \ell 
\le 1$ stability of persistent currents is possible for $\ell = {\rm max}
(x_A, x_B)$, with $x_i = N_i/N$. In what follows below we assume that $x_A >
x_B$.

According to Ref.\,\cite{prl}, the energy spectrum consists of a periodic part, 
plus an envelope parabolic function of $\ell$, in analogy with the problem of 
a single component, as shown by Felix Bloch \cite{FB}. It is thus natural to 
examine the stability of persistent currents at the corresponding values of 
$\ell$ which are higher than unity, i.e., at $\ell = m + x_A$, with $m = 1, 2, 
\dots$ It turns out that for these values of $\ell$ the persistent currents are 
very fragile, even for very small concentrations of the minority component 
\cite{prl}.

Motivated by Ref.\,\cite{prl}, two recent papers by Anoshkin, Wu, and Zaremba 
\cite{p1} and by Wu and Zaremba \cite{p2} examined the same problem theoretically. 
In Ref.\,\cite{p2} it was shown that while for $\ell = m + x_A$, with $m = 1, 2, 
\dots$ persistent currents are absent, indeed, still they are possible for $\ell
= m x_A$. In the language of the present study they investigated the stability
of persistent currents around the combination $(\Psi_A, \Psi_B) = (\Phi_m, \Phi_0)$, 
where $\Phi_m = e^{i m \theta}/\sqrt{2 \pi R}$, with $\theta$ being the azimuthal 
angle and $m = 0, \pm 1, \pm 2, \dots$ To do this, the authors of Ref.\,\cite{p2} 
found solutions of the coupled nonlinear equations satisfied by the two order 
parameters and evaluated their energy. Interestingly, the problem considered here 
may also be viewed from the point of view of solitary-wave solutions, where bound 
states of ``gray" and ``bright" solitary waves in the two components propagate 
together along the torus/annulus \cite{jsmk}. 

In the theoretical analysis of Refs.\,\cite{prl} and \cite{p2} the parameters 
$\gamma_{AA}$, $\gamma_{BB}$, and $\gamma_{AB}$ which characterize the coupling 
between the species $AA$, $BB$, and $AB$ respectively, were assumed to be equal 
to each other. Here $\gamma_{ij} = 4 N a_{ij} R/S$, where $a_{ij}$ ($i = A,B$) 
is the scattering length for elastic atom-atom collisions (assumed to be positive), 
$S$ is the cross section of the toroidal/annular potential, which is approximated 
as a zero-width, ring potential of radius $R$. In our analysis which follows 
below we assume equal values for the couplings $\gamma_{ij}$ and equal masses, $M_A 
= M_B = M$. Under these assumptions the condition for phase coexistence (which is 
crucial in our analysis) is satisfied \cite{prl}. We stress that the problem of 
persistent currents in the case of unequal couplings and/or unequal masses has a 
very different behavior, as we will show in a future publication.

In this study we examine the problem of stability of persistent currents, 
starting with the states $(\Psi_A, \Psi_B) = (\Phi_m, \Phi_n)$. Clearly this pair 
of states has an angular momentum $\ell = m x_A + n x_B$. The benefit from the 
present study is that it provides an alternative solution of the one given in 
Ref.\,\cite{p2}, as it avoids solving the two coupled nonlinear differential 
equations satisfied by $\Psi_A$ and $\Psi_B$. As a result, this procedure is 
also physically transparent. As we explain in more detail below, the combination 
$(\Psi_A, \Psi_B) = (\Phi_m, \Phi_n)$ is not necessarily the lowest-energy (yrast) 
state for the specific value of $\ell = m x_A + n x_B$, and thus one has to 
investigate this problem, too.

In what follows we first investigate in Sec.\,II the question of persistent currents
for $\ell \to (m x_A + n x_B)^-$, assuming that the pair of states $(\Psi_A, \Psi_B) = 
(\Phi_m, \Phi_n)$ constitute the yrast state for the specific value of $\ell$. It turns
out that only the case $n=0$ may give stability of the currents \cite{p2}. Then, in 
Sec.\,III we investigate the conditions which make the pair $(\Psi_A, \Psi_B) = (\Phi_m, 
\Phi_0)$ the actual yrast state. We thus derive two phase boundaries in the plane diagram 
which involves the variables $x_A$ and $\gamma$. In Sec.\,IV we present our numerical 
results. In Sec.\,V we examine the lowest-energy state of the system in the limit of
large values of the coupling, showing that the total density is homogeneous for all
values of the angular momentum. Finally, in Sec.\,VI we give a summary and a discussion 
of our results.

\section{Stability of persistent currents for $\ell \to (m x_A + n x_B)^-$}
 
The Hamiltonian of the system that we consider is 
\begin{eqnarray}
   H = - \sum_{i=1}^{N_A} \frac {\hbar^2} {2 M R^2}  
   \frac {\partial^2} {\partial \theta_{A,i}^2} -
\sum_{j=1}^{N_B} \frac {\hbar^2} {2 M R^2}  \frac 
{\partial^2} {\partial \theta_{B,j}^2}
 \nonumber \\
 + \frac 1 2 U_{AA} \sum_{i \neq j=1}^{N_A} \delta(\theta_{A,i} - \theta_{A,j})
+ \frac 1 2 U_{BB} \sum_{i \neq j=1}^{N_B} \delta(\theta_{B,i} - \theta_{B,j})
\nonumber \\
+ U_{AB} \sum_{i=1}^{N_A} \sum_{j=1}^{N_B} \delta(\theta_{A,i} - \theta_{B,j}).
\end{eqnarray}
Here $U_{ij} = 4 \pi \hbar^2 a_{ij}/(M R S)$ (with $i,j = A,B$), are the matrix elements 
for zero-energy elastic atom-atom collisions. 

To determine the yrast state around the value of the angular momentum $\ell = \ell_0
\equiv m x_A + n x_B$, we first assume that $(\Psi_A, \Psi_B) = (\Phi_m, \Phi_n)$, $m > n 
\ge 0$, is the yrast state for $\ell = \ell_0$; the validity of this assumption is actually 
investigated in Sec.\,III. For values of $\ell \approx \ell_0$, the order parameters 
$\Psi_A$ and $\Psi_B$ will have admixtures of additional states, however while the 
amplitudes of $\Phi_m$, $\Phi_n$ will be of order unity, the amplitudes of these other 
states will be small, much smaller than unity. Furthermore, since we are looking for 
the yrast state, the combination of the (additional) states which will enter $\Psi_A$ 
will be of the form $c_{m-q} \Phi_{m-q} + c_{m} \Phi_{m} + c_{m+q} \Phi_{m+q}$ and 
correspondingly for $\Psi_B$, with $q = 1, 2, \dots$ The reason for this "symmetric" 
choice is that there is a process where two atoms with angular momentum $m$ scatter 
to two other states with angular momentum $m+q$ and $m-q$ (angular momentum is 
conserved in the collisions). The corresponding term in the interaction energy will be 
proportional to $c_{m-q} c_m^2 c_{m+q}$, which may become negative, and thus lower the 
energy. Finally, $q$ is equal to unity, since the states $\Phi_{m-1}$ and $\Phi_{m+1}$ 
have the lowest kinetic energy, while the matrix element of the interaction that is 
associated with the above scattering process is independent of the angular momentum. 
Therefore, we consider the order parameters \cite{prl}
\begin{eqnarray}
\Psi_A &=& c_{m-1} \Phi_{m-1} + c_m \Phi_m + c_{m+1} \Phi_{m+1},
\label{ordpar1}
\\
\Psi_B &=& d_{n-1} \Phi_{n-1} + d_n \Phi_n + d_{n+1} \Phi_{n+1}, 
\label{ordpar2}
\end{eqnarray}
where the six coefficients satisfy the obvious conditions of particle normalization and 
of fixed angular momentum $\ell$. As stated also above, $c_{m \pm 1}$ and $d_{n \pm 1}$ 
are assumed to be small, and thus linearisation is possible. We thus evaluate the expectation
value of the Hamiltonian in the states $\Psi_A$ and $\Psi_B$ and then perform this 
linearisation to find a quadratic expression for the energy per particle, which is
\begin{eqnarray}
 \frac E {N \epsilon} - \frac {\gamma} 2 = 
 x_A [m^2 + (1-2m)c_{m-1}^2 + (1+2m) c_{m+1}^2] 
\nonumber \\
 + x_B [n^2 + (1-2n) d_{n-1}^2 + (1+2n) d_{n+1}^2] 
\nonumber \\
 + \gamma [x_{A}^2 (c_{m-1} + c_{m+1})^2 + x_{B}^2 (d_{n-1} + d_{n+1})^2 
\nonumber \\         
+ 2 x_A x_B (c_{m-1} +c_{m+1})(d_{n-1} + d_{n+1})].
\end{eqnarray}
Here $\epsilon = \hbar^2/(2 M R^2)$ is the kinetic energy and $\gamma = 4 N a R/S$, which was 
defined also above, is the ratio between the interaction energy of the cloud with a homogeneous 
density $N/(2 \pi R)$ of $N = N_A + N_B$ atoms and the kinetic energy $\epsilon$.

The angular momentum $\ell$ in the states of Eqs.\,(\ref{ordpar1}) and (\ref{ordpar2}) is given 
by $\ell = \ell_0 + x_A (c_{m+1}^2 - c_{m-1}^2) + x_B (d_{n+1}^2 - d_{n-1}^2)$. Defining $g = x_A 
(c_{m+1}^2 - c_{m-1}^2) + x_B (d_{n+1}^2 - d_{n-1}^2)$, then $\ell - \ell_0 = g$. Let us thus 
introduce the Lagrange multiplier $\lambda$ and extremize $E/(N \epsilon) + \lambda g$. The 
resulting equations are
\begin{eqnarray}
 -(2m - 1) c_{m-1} + \gamma [x_A (c_{m-1} + c_{m+1}) + x_B (d_{n-1} + d_{n+1})]
\nonumber \\ - \lambda c_{m-1} = 0
 \nonumber \\
 (2m + 1) c_{m+1} + \gamma [x_A (c_{m-1} + c_{m+1}) + x_B (d_{n-1} + d_{n+1})]
\nonumber \\ + \lambda c_{m+1} = 0
 \nonumber \\
 -(2n - 1) d_{n-1} + \gamma [x_A (c_{m-1} + c_{m+1}) + x_B (d_{n-1} + d_{n+1})]
\nonumber \\ - \lambda d_{n-1} = 0
 \nonumber \\
 (2n + 1) d_{n+1} + \gamma [x_A (c_{m-1} + c_{m+1}) + x_B (d_{n-1} + d_{n+1})]
\nonumber \\ + \lambda d_{n+1} = 0.
 \nonumber \\
\label{fbf}
\end{eqnarray}
Demanding that the determinant of the above homogeneous linear system of equations to 
vanish (so that there are non-zero solutions) we get that $\gamma = f(\lambda)$, where
\begin{eqnarray}
 f(\lambda) \equiv \frac 1 2 \frac {[(\lambda + 2m)^2-1] [(\lambda + 2n)^2-1]} 
                       {x_A [(\lambda + 2n)^2-1] + x_B [(\lambda + 2m)^2-1]}.
\label{crit}
\end{eqnarray}
In examining the above condition $\gamma = f(\lambda)$ one has to distinguish between 
the cases $m = n+1$ and $m > n+1$. Starting with the case $m = n+1$, it turns out that
\begin{eqnarray}
 \gamma = f(\lambda) = \frac 1 2 \frac {(\lambda + 2n + 3) [(\lambda + 2n)^2 - 1]} 
                       {\lambda - 1 + 2n + 4 x_B}.
\label{crcon}                       
\end{eqnarray}
The function $f(\lambda)$ has one asymptote, while Eq.\,(\ref{crcon}) has three roots 
for sufficiently large values of $\gamma$. Figure 1(a) shows an example, where we have 
chosen $m = 1$, $n = 0$, $x_A = 0.99$, and $x_B = 0.01$ in this case, while the horizontal
dashed line corresponds to $\gamma = 15$. However, not all three roots are acceptable. In 
order for the angular momentum $\ell$ to be smaller than $\ell_0$ [since we are interested 
in the case $\ell \to \ell_0^-$], only the two larger roots are acceptable. This may be seen 
by examining the sign of 
\begin{eqnarray}
\ell - \ell_0 = x_A (c_{m+1}^2 - c_{m-1}^2) + x_B (d_{n+1}^2 - d_{n-1}^2),
\end{eqnarray}
where, e.g., $c_{m+1}$, and $d_{n \pm 1}$ may be expressed in terms of $c_{m-1}$ from 
Eqs.\,(\ref{fbf}). 

Furthermore, the slope of the energy per particle (or the ``dispersion relation") as 
function of $\ell$ is equal to $-\lambda$. This may be seen from the fact that we have 
extremized $E' = E/(N \epsilon) + \lambda g$ (where $g = \ell - \ell_0$) with respect to 
$\ell$, which implies that $\partial E'/\partial \ell = 0$, or $\partial [E/(N \epsilon)]
/\partial \ell = -\lambda$. As a result, the root that gives the stability is the smaller 
of the two, since we have to choose the one that has the lowest possible energy. Actually, 
this the one close to the asymptote [see the bullet in Fig.\,1(a)], as it was found 
initially in Ref.\,\cite{prl}. This root tends to $\lambda = - 2n + 1 - 4 x_B$ for 
sufficiently large $\gamma$. The slope of the dispersion relation is thus $2n - 1 + 
4 x_B$ and clearly only the case $n = 0$ may give a local energy minimum 
\cite{prl, p1, p2}. 

Turning to the case $m > n+1$, there are two asymptotes and four roots [as in 
Figs.\,1(b) and 1(c), where we have chosen $m=2$ and $n=0$ in the one case and $m=3$ 
and $n=1$ in the other, with $x_A = 0.99$ and $x_B = 0.01$ in both of them]. The dashed 
horizontal line again corresponds to the value of $\gamma = 15$. Again, using the same 
procedure as above, it turns out that among the four possible roots only the higher two 
are acceptable in this case. Again, the root that determines the stability is the one 
with the smaller value, and actually it is the larger of the two roots which result from 
the two asymptotes, indicated as bullets in the plots. For $x_B \to 0$, $\lambda \to -2 n 
+ 1$, or in other words the slope is $2n - 1$. Obviously only the case $n = 0$ may give a 
local energy minimum, again.

From Fig.\,1 it is seen clearly that the root which determines the slope in all three cases 
is positive for sufficiently large values of the coupling \cite{fn} (and thus the slope is 
negative) only in the top and in the middle, but {\it not} in the bottom one. This observation 
is consistent with the fact that only in the case $n=0$ does one get stability of the currents. 

\begin{figure}
\includegraphics[width=8cm,height=5.5cm]{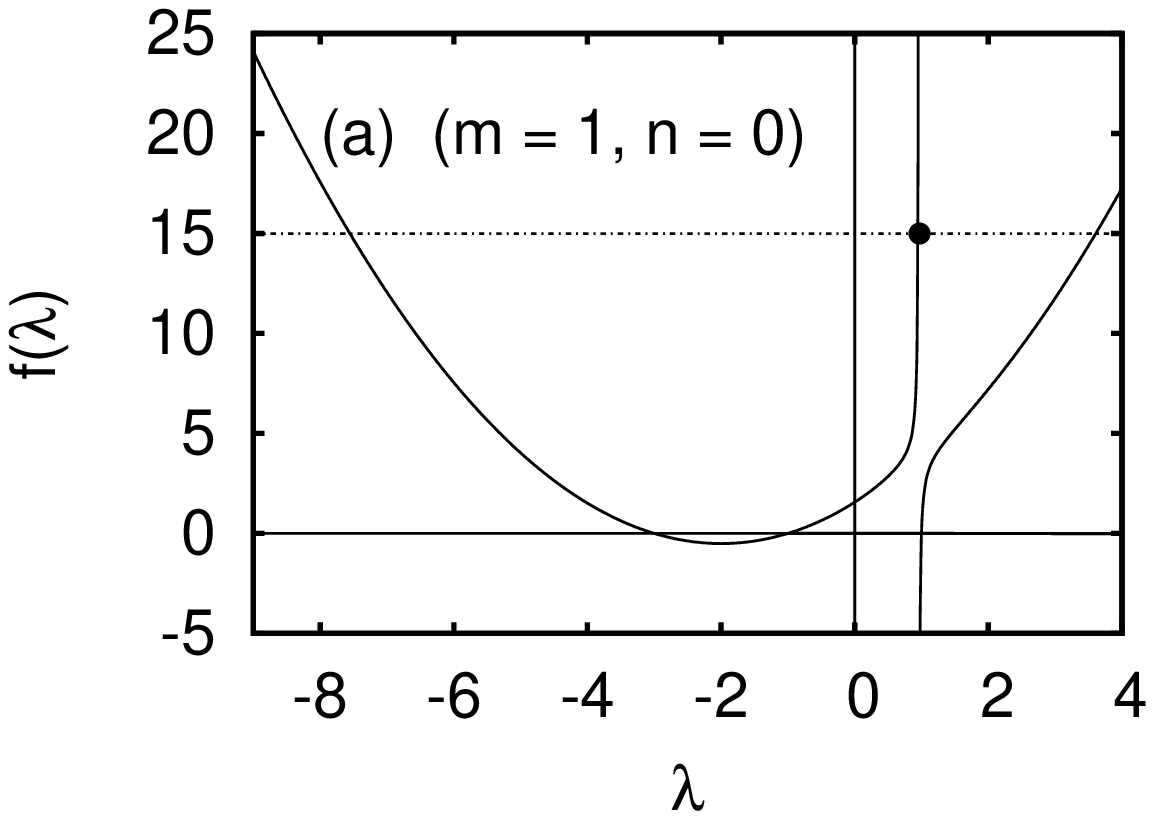}
\includegraphics[width=8cm,height=5.5cm]{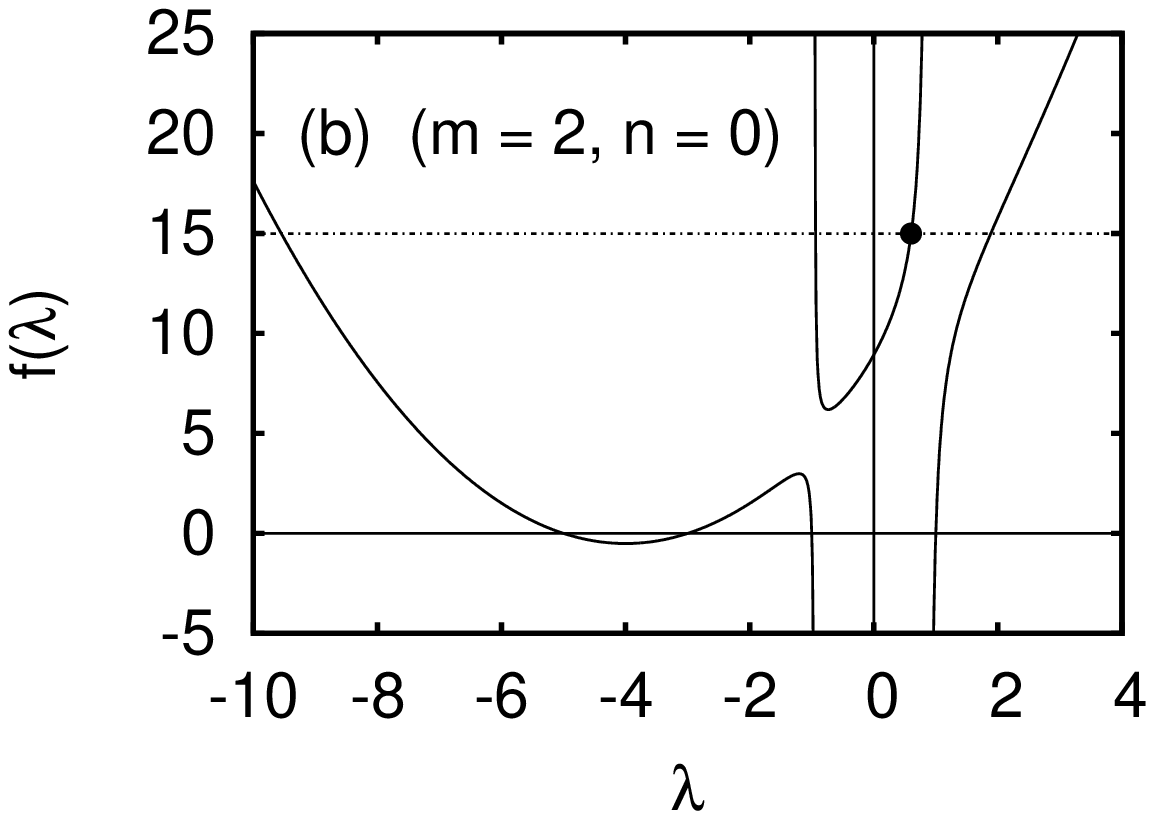}
\includegraphics[width=8cm,height=5.5cm]{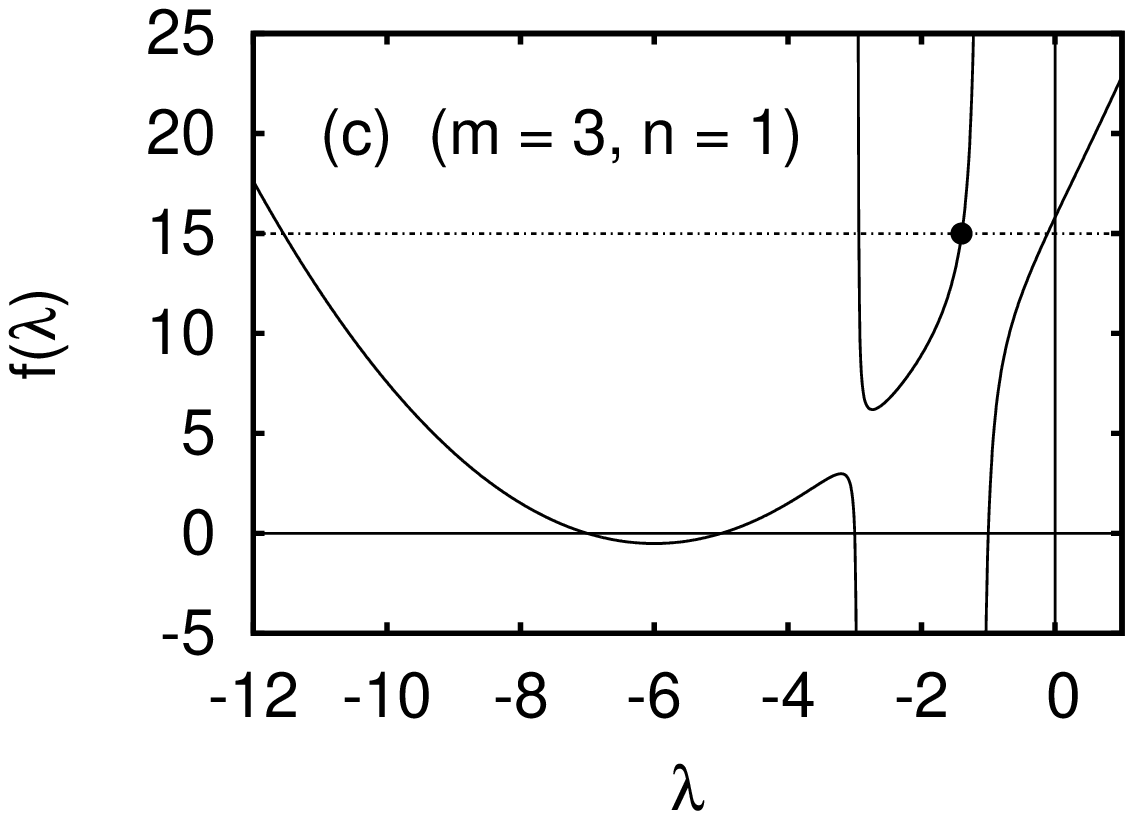}
\caption{The function $f(\lambda)$, for (a) $(\Psi_A, \Psi_B) = (\Phi_1, \Phi_0)$, (b) 
$(\Psi_A, \Psi_B) = (\Phi_2, \Phi_0)$, and (c) $(\Psi_A, \Psi_B) = (\Phi_3, \Phi_1)$.
Here $x_A =0.99$ and $x_B = 0.01$. The horizontal dashed line refers to $\gamma = 15$,
while the bullets show the roots which determine the slope in each case.}
\end{figure} 

To get the critical value of $\gamma$ for stability of the persistent currents we thus 
set $\lambda = 0$ and $n=0$ in Eq.\,(\ref{crit}), getting
\begin{eqnarray}
  \gamma_{\rm cr} = \frac 1 2 
  \frac {4 m^2 - 1} {1 - 4 m^2 x_B}.
\label{me}
\end{eqnarray} 
The above expression has been derived in Ref.\,\cite{p2}, and it has an asymptote at 
\begin{eqnarray}
 x_{A, \rm cr} = 1 - \frac 1 {4 m^2},
\label{asymp} 
\end{eqnarray}
and therefore $x_A$ is bounded from below by this value. From Eq.\,(\ref{me}) it also
follows that for $x_A \to 1$,
\begin{eqnarray} 
     \gamma_{\rm cr} \to 2 m^2 - \frac 1 2,
\label{crg111}
\end{eqnarray}
which is the well-known result for the case of one component (see, e.g., Ref.\,\cite{prl}). 

\section{Yrast state for $\ell = m x_A$}

In the previous section we assumed implicitly that the pair of states $(\Psi_A, 
\Psi_B) = (\Phi_m, \Phi_0)$ gives the yrast state for the specific value of the 
angular momentum $\ell = m x_A$ ($n$ is set equal to zero from now on, since this 
is the only possible value that may give stability of the currents). For $\ell = m, 
m + x_B, m + x_A$, and $m + 1$, where $m = 0, \pm 1, \pm 2, \dots$, the yrast state 
consists of the pairs $(\Psi_A, \Psi_B) = (\Phi_m, \Phi_m)$, $(\Psi_A, \Psi_B) = 
(\Phi_m, \Phi_{m+1})$, $(\Psi_A, \Psi_B) = (\Phi_{m+1}, \Phi_m)$, and $(\Psi_A, 
\Psi_B) = (\Phi_{m+1}, \Phi_{m+1})$, respectively, for any value of $\gamma$. 

For any other value of the angular momentum the yrast state may consist of a
combination of more than one modes for $\Psi_A$ and $\Psi_B$. For example, for 
weak interatomic interactions and $0 \le \ell \le 1$, the whole yrast state 
consists only of the two lowest-energy modes, i.e., 
\begin{eqnarray}
 \Psi_A = c_0 \Phi_0 + c_1 \Phi_1, \,\,\, \Psi_B = d_0 \Phi_0 + d_1 \Phi_1.
\end{eqnarray}
On the other hand, for sufficiently strong values of $\gamma$, states of a homogeneous 
density distribution are candidates for being the yrast states, since they minimize the 
interaction energy. The combination $(\Psi_A, \Psi_B) = (\Phi_m, \Phi_0)$ has a homogeneous 
density distribution (in each component separately), and this pair of states is indeed the
yrast state under the conditions examined below.

To attack this problem one may use the approach described in the previous section, 
however in the present case the Lagrange multiplier is not set equal to zero, but
rather in addition to Eq.\,(\ref{crit}) (with $n=0$) self-consistency of 
Eqs.\,(\ref{fbf}) and the equation for the angular momentum introduces the 
additional condition
\begin{eqnarray}
 4 \gamma^2 x_A x_B (\lambda + 2 m) + [(\lambda + 2 m)^2 - 2 \gamma x_A - 1]^2 
 \lambda  = 0,
\nonumber \\
\label{add}
\end{eqnarray}
where $-2 m < \lambda < 0$. Using Eqs.\,(\ref{crit}) and (\ref{add}) one may eliminate 
$\lambda$ and thus get the critical value of $\gamma$, $\gamma_{\rm cr}^{\rm yr}$, as a 
function of $x_A$ (or equivalently $x_B$). While this has to be done numerically in 
general, the limiting cases may be handled analytically. For $x_A \to 1$, then $\lambda 
\to -1$, and thus 
\begin{eqnarray}
 \gamma_{\rm cr}^{\rm yr} \to 2 m (m - 1),
\label{addd}
\end{eqnarray}
in agreement with the result of Wu and Zaremba \cite{p2}. Comparing the above value of 
$\gamma_{\rm cr}^{\rm yr}$ of Eq.\,(\ref{addd}) with $\gamma_{\rm cr}$, given by 
Eq.\,(\ref{crg111}), it turns out that $\gamma_{\rm cr}^{\rm yr} < \gamma_{\rm cr}$. 

In addition, for large values of $\gamma$, then $\lambda \to -2 m x_B$ and thus
there is one asymptote for the following value of $x_A$
\begin{eqnarray}
   x_{A, \rm{cr}}^{\rm yr} = \frac 1 2 + \frac {\sqrt{m^2 - 1}} {2 m},
\label{asympto}
\end{eqnarray}
where the solution with the negative sign is not acceptable. Again, the above expression 
agrees with the one given by Ref.\,\cite{p2}. Comparing this value of 
$x_{A, \rm{cr}}^{\rm yr}$ from Eq.\,(\ref{asympto}) with $x_{A, \rm cr}$ given by 
Eq.\,(\ref{asymp}) we see that $x_{A, \rm{cr}}^{\rm yr} < x_{A, \rm cr}$. 

\section{Relative position of the two phase boundaries}

In addition to the above analytical results, we investigated numerically (variationally) 
the phase boundary for the combination $(\Psi_A, \Psi_B) = (\Phi_m, \Phi_0)$ to become 
the yrast state (i.e., the question investigated in Sec.\,III). We have thus kept the two 
neighbouring states around $\Phi_m$ and $\Phi_0$, namely $\Phi_{m \pm 1}$ and 
$\Phi_{\pm 1}$. 

The result of this calculation is shown as the dashed curve in Fig.\,2 for $(m,n) = (2,0)$. 
This curve terminates due to numerical reasons. Our results, however, are consistent with the 
divergence of $\gamma_{\rm cr}^{\rm yr}$ for $x_A \to (2 + \sqrt 3)/4 \approx 0.933$, according 
to Eq.\,(\ref{asympto}). Also, for $x_A \to 1$, $\gamma_{\rm cr}^{\rm yr} \to 4$, in agreement 
with Eq.\,(\ref{addd}). 

More generally, the whole phase boundary that defines the pair $(\Psi_A, \Psi_B) = (\Phi_m, \Phi_0)$ 
to be the actual yrast state coincides with the one that results from the analysis presented in 
Sec.\,III, i.e., from the solution of Eqs.\,(\ref{crit}) and (\ref{add}). In the same figure 
we have also included the phase boundary for the stability of persistent currents for $(m,n) 
= (2,0)$, Eq.\,(\ref{me}), [again for $(m,n) = (2,0)$], where $\gamma_{\rm cr} =  7.5/(16 x_A - 15)$.

Thus, the results of this section confirm the general picture which implies that the phase for 
stability of the currents is always included within the phase for the pair $(\Psi_A, \Psi_B) = 
(\Phi_m, \Phi_0)$ to be the yrast state, and thus the results of Sec.\,II -- which rely on this 
crucial assumption -- are always valid.

\begin{figure}
\includegraphics[width=8cm,height=5.5cm]{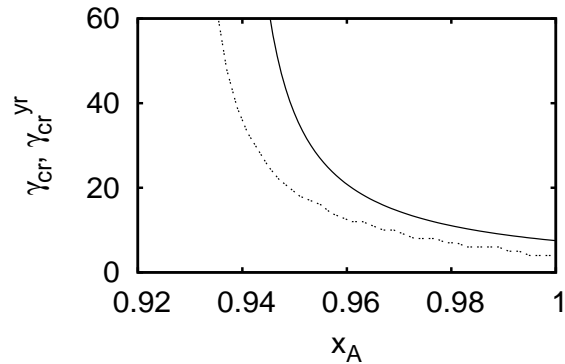}
\caption{The two phase boundaries showing $\gamma_{\rm cr}^{\rm yr}$ (dashed curve) 
and $\gamma_{\rm cr}$ (solid curve) versus $x_A$, for $(\Psi_A, \Psi_B) = (\Phi_2, \Phi_0)$.}
\end{figure} 

\section{A unified description of the yrast state for large values of the coupling}

One remarkable result of the analysis presented in Sec.\,II is that the slope of the 
solution that determines the stability of the currents saturates, as it has an upper 
bound, even for large values of $\gamma$. This is easily seen, since the solution that 
determines the stability comes from either the only asymptote in the case $(m=1, n=0)$,
or one of the two asymptotes for ($m>1, n=0$) of $f(\lambda)$. More specifically, 
for $x_B \to 0$, the slope of the dispersion relation for large values of $\gamma$ 
tends to $-1 + 4 x_B$ when $m=1$, and it tends to $-1 + 2 m (m-1) x_B$ when $m > 1$. 

To get some insight into this result, let us examine the density, which is given by
(since $|c_{m \pm 1}| \ll |c_m|$ and $|d_{\pm 1}| \ll |d_0|$),
\begin{eqnarray}
 n_A(\theta) &=& \frac {x_A} {2 \pi R} [1 + 2(c_{m-1}+c_{m+1}) \cos \theta],
 \nonumber \\
 n_B(\theta) &=& \frac {x_B} {2 \pi R} [1 + 2(d_{-1}+d_{1}) \cos \theta].
\end{eqnarray}
Examining the term $x_A (c_{m-1}+c_{m+1}) + x_B (d_{-1}+d_{1})$ that appears in the 
total density $n_A + n_B$, it turns out that this is proportional to
\begin{eqnarray}
  \frac {x_A} {(\lambda + 2 m)^2-1} + \frac {x_B} {\lambda^2 - 1} \propto \gamma^{-1} 
  \to 0,
\end{eqnarray}  
as Eq.\,(\ref{crit}) implies. In other words, the density variation of $n_A + n_B$ is 
constant (to order $1/\gamma$), which is the reason for the saturation of the slope: 
the system manages to maintain its density homogeneous (thus gaining potential energy) 
at the expense of kinetic energy (due to the extra components $\Phi_{m \pm 1}$ and 
$\Phi_{\pm 1}$ in the order parameters $\Psi_A$ and $\Psi_B$, respectively). For example,
for $m=1$ it may be seen after some algebra that for $\ell \to x_A^-$
\begin{eqnarray}
  \frac E {N \epsilon} - \frac {\gamma} 2 = \ell + 2 (1 - 2 x_B) (x_A - \ell). 
\end{eqnarray}
From this formula it follows that the difference $E/{N \epsilon} - {\gamma}/2$ is 
indeed due to the kinetic energy. The same equation also implies that the slope is 
$-1 + 4 x_B$, as we argued also above.

Actually, the homogeneity of the total density is a more general result, which 
characterizes the yrast state for all values of the angular momentum when the coupling 
is sufficiently large. As shown in Ref.\,\cite{prl}, for $0 \le \ell \le x_B$ and $x_A 
\le \ell \le 1$, the total density is homogeneous exactly for any value of $\gamma$. 
The same result holds for large values of $\gamma$, also for $x_B \le \ell  \le x_A$, 
and thus for {\it all} values of $\ell$. 

To demonstrate this, we performed a constrained minimization of the energy of the system 
considering the trial order parameters $\Psi_A = \sum_{m=-1}^{m=3} c_m \Phi_m$ and $\Psi_B 
= \sum_{m=-2}^{m=2} d_m \Phi_m$. The imposed constraints were that of particle 
normalization, of a fixed angular momentum, and finally a constant total density 
distribution. Figure 3 shows the result of this calculation. In this calculation the
coupling drops out completely from the energy, apart from the ``background" energy of 
the homogeneous density distribution [which has an energy $E/(N\epsilon) = \gamma/2$]. 
This calculation thus demonstrates the saturation that we described earlier. The homogeneity 
of the total density distribution is expected to become asymptotically exact for large 
values of $\gamma$. In a sense, this effect is analogous to the fermionization of hard-core 
bosons in one dimension, where the system pays kinetic energy via the fermionization of the 
bosons, however it gains (more) energy because of the assumed large value of the coupling 
between the particles. 

In the same figure we also show the result of the minimization of the energy via the 
method of imaginary-time propagation, for some fixed and relatively large value of the 
coupling $\gamma = 1250/\pi^2$. In this calculation, in order to fix the expectation value 
of the angular momentum, we have used a Lyapunov functional \cite{KCP}. While close to
each other, the two curves do not coincide, for three reasons. The first one is that they 
do not correspond to the same value of $\gamma$. The second reason is that the one for 
finite $\gamma$ is exact up to numerical error. The third reason is that the one for finite 
$\gamma$ does not have the constraint of an exactly homogeneous density distribution. As 
we saw earlier, variations in the total density of order $1/\gamma$ are expected to be 
present, as we have also confirmed in the numerical solution we have found with the method 
of the imaginary-time propagation.

Further evidence for the homogeneity of the total density distribution is also shown in 
Fig.\,4, where we plot the density of the two species $n_A(\theta)$, $n_B(\theta)$, and 
also the total density distribution $n_A + n_B$, for the values used in Fig.\,3 ($x_A = 
0.8$, $x_B = 0.2$, and $\gamma = 1250/\pi^2$) and also choosing $\ell$ to be 0.3. While 
the density of the two components shows a substantial variation over the ring, the total 
density is very close to homogeneous, with fluctuations which are of order $1/\gamma$, as 
we have checked from our data.

\begin{figure}
\includegraphics[width=8cm,height=6cm]{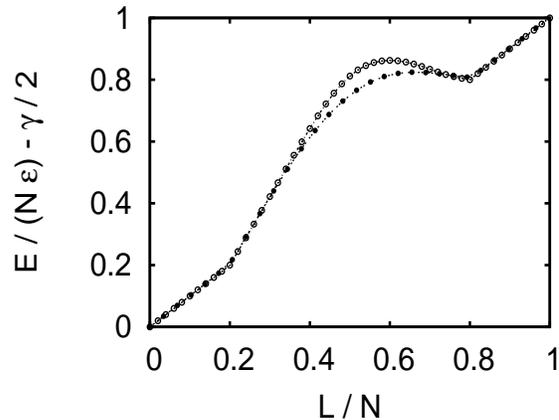}
\caption{The dispersion relation for $x_A = 0.8$ and $x_B = 0.2$, with $\gamma = 
1250/\pi^2$ in the lower curve and ``large" $\gamma$ in the higher, evaluated through 
the two methods described in Sec.\,V.}
\end{figure} 

\begin{figure}
\includegraphics[width=8cm,height=6cm]{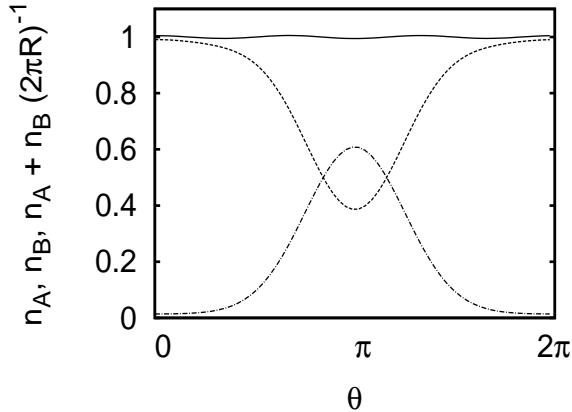}
\caption{The density $n_A(\theta)$ (dotted curve), $n_B(\theta)$ (dashed dotted curve), 
and the total density $n_A(\theta) + n_B(\theta)$ (solid curve), for $\ell = 0.3$, $x_A 
= 0.8$, $x_B = 0.2$, and $\gamma = 1250/\pi^2$ that results from the method of imaginary 
time propagation described in Sec.\,V.}
\end{figure} 

\section{Discussion and conclusions}

In the present study we have examined the problem of stability of persistent 
currents of a mixture of two Bose-Einstein condensates which are confined in a 
ring potential considering the combination $(\Psi_A, \Psi_B) = (\Phi_m, \Phi_n)$.
We have thus found that only the case $(\Psi_A, \Psi_B) = (\Phi_m, \Phi_0)$ may give 
rise to stability of the currents, in agreement with the results of Ref.\,\cite{p2}. 
The (essentially variational) method that we have used gives insight into this problem, 
since it avoids solving the two coupled nonlinear differential equations. On the other 
hand, there is an additional complication, which has to do with the states $(\Psi_A, 
\Psi_B) = (\Phi_m, \Phi_0)$ being the actual yrast states. 

As we have seen, for sufficiently strong interactions and a large population imbalance 
this combination becomes the yrast state, as seen in the lower curve of Fig.\,2, for 
$m=2$. For even larger values of the coupling the combination $(\Psi_A, \Psi_B) = 
(\Phi_m, \Phi_0)$ becomes a local minimum of the dispersion relation, provided that 
the above solution does not belong to the linear part of the dispersion \cite{prl}. 

In order for our analysis of the stability of the persistent currents to be valid -- 
which investigates the behavior of the system at $\ell \to (m x_A)^-$ -- requires 
that the combination $(\Psi_A, \Psi_B) = (\Phi_m, \Phi_0)$ is the yrast state for 
$\ell = m x_A$. In the ($x_A$ -- $\gamma$) phase diagram, there are thus two phase 
boundaries which need to be derived. It turns out that for $n = 0$ the condition for 
stability of the currents guarantees that the pair $(\Psi_A, \Psi_B) = (\Phi_m, \Phi_0)$ 
is the yrast state [for $\ell = m x_A$ and $x_A > x_{A, \rm cr} = 1 - 1/(4 m^2)$]. As 
seen from Fig.\,2, for some fixed population imbalance and some fixed angular momentum 
$\ell = m x_A$, as the coupling increases, first the combination $(\Psi_A, \Psi_B) = 
(\Phi_m, \Phi_0)$ becomes the yrast state and then it provides a local minimum in the 
dispersion relation.

Another remarkable and general result of our study is the homogeneity of the total density 
distribution (see Fig.\,4), which characterizes the {\it whole} yrast spectrum for large 
values of the coupling constant. In this limit the yrast state has this surprisingly simple 
feature, in a sense resembling the Tonks-Girardeau limit of fermionized bosons.

Last but not least, it is worth comparing the above results (which assume that 
$\gamma_{AA} = \gamma_{AB} = \gamma_{BB}$) with the case where these are unequal, 
which will be examined in a future publication. The main difference (in terms 
of the applicability) of the main result of this study, i.e., Eq.\,(\ref{me}), is 
when $x_B$ is not sufficiently small. For equal values of the $\gamma_{ij}$ considered 
here, stability of persistent currents is not possible. On the other hand, for unequal 
values of the $\gamma_{ij}$ metastability may be possible for sufficiently strong
interatomic interactions. In addition, the case $n > 0$ may also give rise to 
persistent currents, as opposed to the present problem. Finally, more local minima
may appear in the dispersion relation.

Given that an experiment on this problem has already been performed \cite{Zoran}, it
would be interesting to investigate whether the interesting structure revealed in 
the theoretical studies of this problem is indeed observable.

\acknowledgements

We thank Zhigang Wu and Eugene Zaremba for pointing out that the case $n > 0$
requires special care. This project is implemented through the Operational Program 
"Education and Lifelong Learning", Action Archimedes III and is co-financed by the 
European Union (European Social Fund) and Greek national funds (National Strategic 
Reference Framework 2007 - 2013).


\begin{thebibliography}{99}
\bibitem{fn} We stress here that when we refer to ``large" values of the coupling we restrict 
ourselves to the case where, although $\gamma$ is large, still $\gamma \ll N^2$, so that the 
gas is still in the mean-field limit, away from the Tonks-Girardeau regime.
\end{thebibliography}
\end{document}